\documentclass[sigconf]{acmart}
\usepackage{algorithmic}
\usepackage{graphicx}
\usepackage{textcomp}
\usepackage{xcolor}
\usepackage{url}
\usepackage{booktabs}
\usepackage{tcolorbox}
\usepackage{subcaption}
\usepackage{balance}

\usepackage{listingsutf8}
\lstdefinelanguage{docker}{
  keywords={FROM, RUN, COPY, ADD, ENTRYPOINT, CMD,  ENV, ARG, WORKDIR, EXPOSE, LABEL, USER, VOLUME, STOPSIGNAL, ONBUILD, MAINTAINER},
  keywordstyle=\color{blue}\bfseries,
  identifierstyle=\color{black},
  sensitive=false,
  comment=[l]{\#},
  commentstyle=\color{purple}\ttfamily,
  stringstyle=\color{red}\ttfamily,
  morestring=[b]',
  morestring=[b]"
}
\lstdefinelanguage{docker-compose}{
  keywords={image, environment, ports, container_name, ports, volumes, links},
  keywordstyle=\color{blue}\bfseries,
  identifierstyle=\color{black},
  sensitive=false,
  comment=[l]{\#},
  commentstyle=\color{purple}\ttfamily,
  stringstyle=\color{red}\ttfamily,
  morestring=[b]',
  morestring=[b]"
}
\lstdefinelanguage{docker-compose-2}{
  keywords={version, volumes, services},
  keywordstyle=\color{blue}\bfseries,
  keywords=[2]{image, environment, ports, container_name, ports, links, build},
  keywordstyle=[2]\color{olive}\bfseries,
  identifierstyle=\color{black},
  sensitive=false,
  comment=[l]{\#},
  commentstyle=\color{purple}\ttfamily,
  stringstyle=\color{red}\ttfamily,
  morestring=[b]',
  morestring=[b]"
}
\begin{document}

\title{
Meta-Maintanance for Dockerfiles: Are We There Yet?}

\newcommand{\RqOne}{\textbf{RQ1:} \emph{How Dockerfiles with version-pinned packages are maintained?}}

\newcommand{\RqTwo}{\textbf{RQ2:} \emph{What package sets present?}}

\newcommand{\RqThree}{\textbf{RQ3:} \emph{How do package sets evolve in repository groups?}}

\newcommand{\RqFour}{\textbf{RQ4:} \emph{How do developers consider the proposed package set update?}}

\graphicspath{{figs}}

\author{Takeru Tanaka}
\affiliation{%
  \institution{Nara Institute of Science and Technology}
  \country{Japan}
}
\email{tanaka.takeru.tr5@is.naist.jp}

\author{Hideaki Hata}
\affiliation{%
  \institution{Shinshu University}
  \country{Japan}
}
\email{hata@shinshu-u.ac.jp}

\author{Bodin Chinthanet}
\affiliation{%
  \institution{Nara Institute of Science and Technology}
  \country{Japan}
}
\email{bodin.ch@is.naist.jp}

\author{Raula Gaikovina Kula}
\affiliation{%
  \institution{Nara Institute of Science and Technology}
  \country{Japan}
}
\email{raula-k@is.naist.jp}

\author{Kenichi Matsumoto}
\affiliation{%
  \institution{Nara Institute of Science and Technology}
  \country{Japan}
}
\email{matumoto@is.naist.jp}

\renewcommand{\shortauthors}{Tanaka et al.}

\begin{abstract}
Docker allows for the packaging of applications and dependencies, and its instructions are described in Dockerfiles.
Nowadays, version pinning is recommended to avoid unexpected changes in the latest version of a package.
However, version pinning in Dockerfiles is not yet fully realized (only 17k of the 141k Dockerfiles we analyzed), because of the difficulties caused by version pinning.
To maintain Dockerfiles with version-pinned packages, it is important to update package versions, not only for improved functionality, but also for software supply chain security, as packages are changed to address vulnerabilities and bug fixes.
However, when updating multiple version-pinned packages, it is necessary to understand the dependencies between packages and ensure version compatibility, which is not easy.
To address this issue, we explore the applicability of the meta-maintenance approach, which aims to distribute the successful updates in a part of a group that independently maintains a common artifact.
We conduct an exploratory analysis of 7,914 repositories on GitHub that hold Dockerfiles, which retrieve packages on GitHub by URLs.
There were 385 repository groups with the same multiple package combinations, and 208 groups had Dockerfiles with newer version combinations compared to others, which are considered meta-maintenance applicable.
Our findings support the potential of meta-maintenance for updating multiple version-pinned packages and also reveal future challenges.

\end{abstract}





\maketitle

\section{Introduction}

Docker is a widely used containerization tool that allows developers to avoid the problem of \textit{dependency hell} by packaging applications and dependencies for each environment and running them in an isolated execution environment~\cite{merkel2014docker}.
However, previous empirical studies on Dockerfiles that describe the steps to build containers have revealed that dependency issues occur frequently~\cite{Cito:MSR2017:1,Henkel:ICSE2021:25}.

Pinning to a specific version of a package is called \textit{version pinning} and is recommended to reproduce the expected environment when building containers at different times~\cite{10.1371/journal.pcbi.1008316}. Version pinning in a Dockerfile is expected to avoid build failures due to unexpected changes in the latest version of packages.
Not limited to Dockerfiles, version pinning is a common way to ensure fine-grained control over software packages. This practice reduces the risk of changes, unknown bugs, and vulnerabilities that may only be present in new releases~\cite{9519532}.
However, fixing dependencies prevents the adoption of important security and bug fixes. Thus, unless constantly updated to new versions of the constraints, projects will become less secure over time~\cite{9519532}. This problem manifests itself when Dockerfiles are used for long periods of time.
Previous studies on Dockerfiles have shown that package version updates are insufficient, despite the fact that containers contain many old and vulnerable packages~\cite{Shu:CODASPY2017:24,Liu:ESORICS2020}.
Therefore, periodic version updates are essential, and the problem of dependencies reappears in Dockerfiles.

As seen in the recent Apache Log4j vulnerability issue,\footnote{\url{https://logging.apache.org/log4j/2.x/security.html}} software supply chain security has become essential and there are several social movements.
To improve and secure open source software supply chain issues, the Open Security Software Foundation (OpenSSF), a project of the Linux Foundation, has announced the launch of the Alpha-Omega Project.\footnote{https://openssf.org/community/alpha-omega/}
This project aims to improve the security of the global open source software (OSS) supply chain by working with project maintainers to systematically find and fix new vulnerabilities in open source code.
The ``Alpha'' supports the most important OSS projects, while the ``Omega'' targets at least 10,000 widely deployed OSS projects.
GitHub has a number of blog articles tagged ``supply chain security'' from 2020,\footnote{\url{https://github.blog/tag/supply-chain-security/}}
offers a service called Dependabot that checks for new versions of dependencies and submits pull requests for security updates to fix vulnerable dependencies.\footnote{\url{https://docs.github.com/en/code-security/dependabot}}
Dependabot supports updating tags in Dockerfile base images. Although support for package version updates has been requested by users on a GitHub issue, it has not yet been implemented~\cite{Web:GHissue}.


Version updates for software supply chain security have been highlighted as an important challenge, but there is still not enough support for Dockerfiles.
In particular, to update the versions of multiple version-pinned packages in a Dockerfile, it is necessary to verify whether there are dependencies between the packages and whether they satisfy version compatibility, but to our knowledge there are no tools that address this issue.

Meta-maintenance is a unique approach to the dependency problem that seeks to support multiple projects rather than individual projects by distributing existing advanced outcomes.
In detail, meta-maintenance identifies useful changes from a subset of a repository group that maintains the same artifacts and sharing them with other projects~\cite{Hata:ICSE2021}.

This paper explores the applicability of meta-maintenance to the problem of updating multiple version-pinned packages.
In addition, we believe that meta-maintainance is an important step toward making the software supply chain more secure.
Our objective is to understand the how meta-maintainance can be applied to updating Dockerfiles.
We report on the results of an exploratory study of 7,914 repositories on GitHub that hold Dockerfiles with version-pinned packages.
We find that: only 18\% of the repositories had updated version-pinned packages in their Dockerfiles,
multiple-package combinations appearing in multiple repositories consist of a minimum of two and a maximum of 38 packages,
repository groups to which more than half of the package sets belong contain Dockerfiles with more recent version combinations compared to other Dockerfile version combinations, which are candidates for meta-maintenance.
From the responses to our submitted pull requests, we have learned that developers may need additional motivation to consider updates.
This study provides insights into the characteristics and challenges of Dockerfile meta-maintenance.


\section{Dockerfile Meta-Maintenance}
\label{sec:terminology}

This section describes version-pinned packages to be analyzed in this paper and the corresponding idea for the meta-maintenance approach.

\subsection{Version-Pinned Packages}

There are two main ways to install packages in a Dockerfile.
The first method is to use package management systems. As shown in the example below, the latest versions of packages are installed by passing package names as arguments to the \texttt{apt-get} command.
\begin{lstlisting}[language=docker,breaklines=true]
RUN apt-get update && apt-get -y install git
\end{lstlisting}

The second method is to use the \texttt{curl} or \texttt{wget} commands in a Dockerfile to directly obtain packages.
When using these commands, the URLs of the servers from which the packages are obtained are specified as arguments to the \texttt{curl} or \texttt{wget} commands, as shown in the example below.\footnote{This example can be found in the ``Best practices for writing Dockerfiles'' section of the Docker Documentation. \url{https://docs.docker.com/develop/develop-images/dockerfile_best-practices/}}
\begin{lstlisting}[language=docker,breaklines=true]
ENV PG_VERSION=9.3.4
    RUN curl -SL https://example.com/postgres-$PG_VERSION.tar.xz | tar -xJC /usr/src/postgres && ...
\end{lstlisting}

The second case seems to require more support as developers have to resolve the dependencies of multiple packages themselves. In addition, Dependabot developers noted difficulty in properly parsing the first case~\cite{Web:GHissue}.
Therefore, this paper only targets the second case.
If package versions are pinned, URLs will be different for each release, thus developers must change URLs in Dockerfiles to update package versions.
However, updating multiple package versions is costly because of the need to verify whether there are dependencies between packages and whether version compatibility is met.

\subsection{Meta-Maintenance Approach}

Toward supporting multiple version-pinned packages updates, this paper explores the applicability of the meta-maintenance approach, which aims to distribute advanced outcomes from a portion of a group that independently maintains common artifacts to the entire group~\cite{Hata:ICSE2021}.
The idea is to find Dockerfiles that use a newer version combination within a group of Dockerfiles that install the same multiple packages, and to update all Dockerfiles together by suggesting that version combination to the other Dockerfiles in the group.
If some repositories adopt a newer version combination, that combination is considered a useful combination that could satisfy version compatibility.

The original study defined a \textbf{seed} file as a specific version of a file shared by multiple software repositories, and software repositories containing the same seed file as a \textbf{seed family}~\cite{Hata:ICSE2021}.
For Dockerfile meta-maintenance, we consider a combination of multiple packages (a \textbf{package set}) is considered a \textbf{seed}, and software repositories with Dockerfiles containing that package set is considered a \textbf{seed family}.
While file-level meta-maintenance analyzed how files belonging to the same families evolve, this paper explores new version combinations to package sets in the same families.

\section{Research Questions}

The main goal of this study is to explore the applicability 
of the meta-maintenance approach for updating multiple version-pinned packages in Dockerfiles.
Based on this goal, we construct four research questions to guide our study.
We now present each of these questions, along with our motivation.

\noindent
\RqOne~

\noindent
\RqTwo~

\noindent
\RqThree~

\noindent
\RqFour~

First, \textbf{RQ1} investigates how much Dockerfiles, installing version-pinned packages, have been updated to understand if there is a need to update packages and there are any changes that could be useful in meta-maintenance. We analyze the presence of updates for each Dockerfile and each repository.
The motivation for \textbf{RQ2} is to understand what package combinations (package sets) appear in Dockerfiles in multiple repositories. We show the distributions of the number of packages contained in a package set and the number of repositories with same package sets. Also we investigate the programming languages in which the packages included in the package set are written.
\textbf{RQ3} then explores the package sets from an evolutionary and maintenance perspective. We would like to understand how Dockerfiles with the same package set evolve differently.
Our aim for \textbf{RQ4} is to understand how developers react to proposed package updates in their projects.



\begin{figure*}
    \centering
    \includegraphics[width=\linewidth]{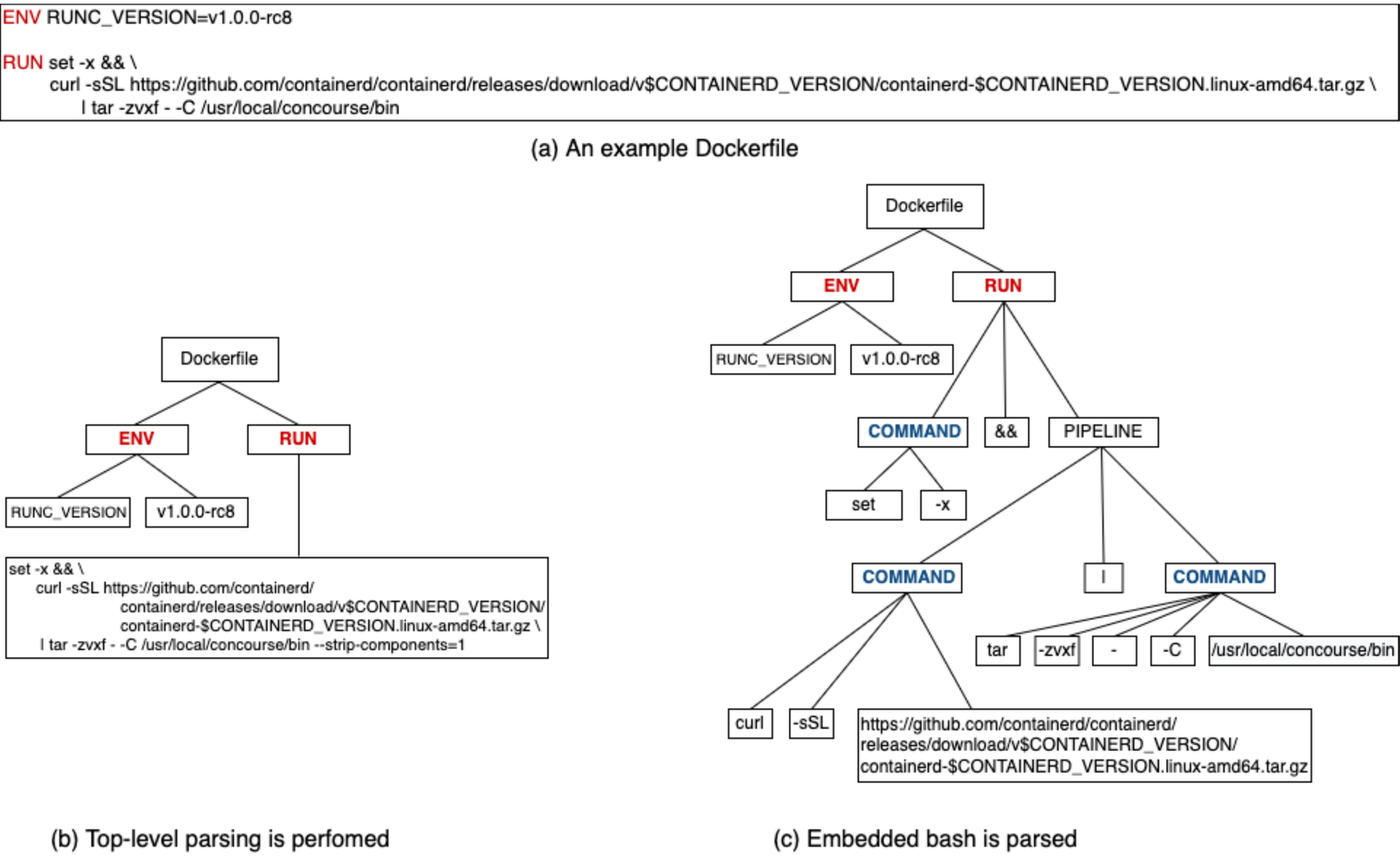}
	\caption{An example Dockerfile at each phase of parsing.}
    \label{fig:parse}
\end{figure*}

\section{Data Collection}

This section describes the procedure for preparing repositories, obtaining Dockerfiles, extracting URLs, and identifying version-pinned packages.

\subsection{Preparing Repositories}
Since the list of repositories that hold Dockerifiles cannot be retrieved directly from GitHub,
we collected repositories that were created between 2008 and 2020, had at least 10 stars when accessed, and were not forks of other repositories.
A list of repositories that fit the criteria was created in January 2021 using the GitHub API v3,\footnote{\url{https://developer.github.com/v3/}} and repositories were collected from January to March 2021, resulting in 1,210,272 repositories collected.

\subsection{Obtaining Dockerfiles}
\label{sec:obtaining-dockerfiles}
Since `$*.Dockerfile$' and `$Dockerfile.*$' are possible Dockerfile names, we searched for files with the pattern `$.*(d\verb/|/D)ockerfile.*$' on the HEAD of the default branch of each repository we obtained.
As a result, we identified 463,465 candidate Dockerfiles. 
To remove invalid files as Dockerfile from the obtained Dockerfile candidates, the files were examined using a Dockerfile parser.\footnote{\url{https://github.com/moby/buildkit/tree/master/frontend/dockerfile/parser}} Dockerfile must consist of one or more instructions and start with a FROM instruction that specifies a base image. Therefore, we analyzed each candidate Dockerfile and obtained a valid Dockerfile if it was parsable and started with the FROM instruction. As a result, 77,469 repositories holding 434,846 Dockerfiles were identified.

\begin{figure*}
    \centering
    \includegraphics[width=.9\linewidth]{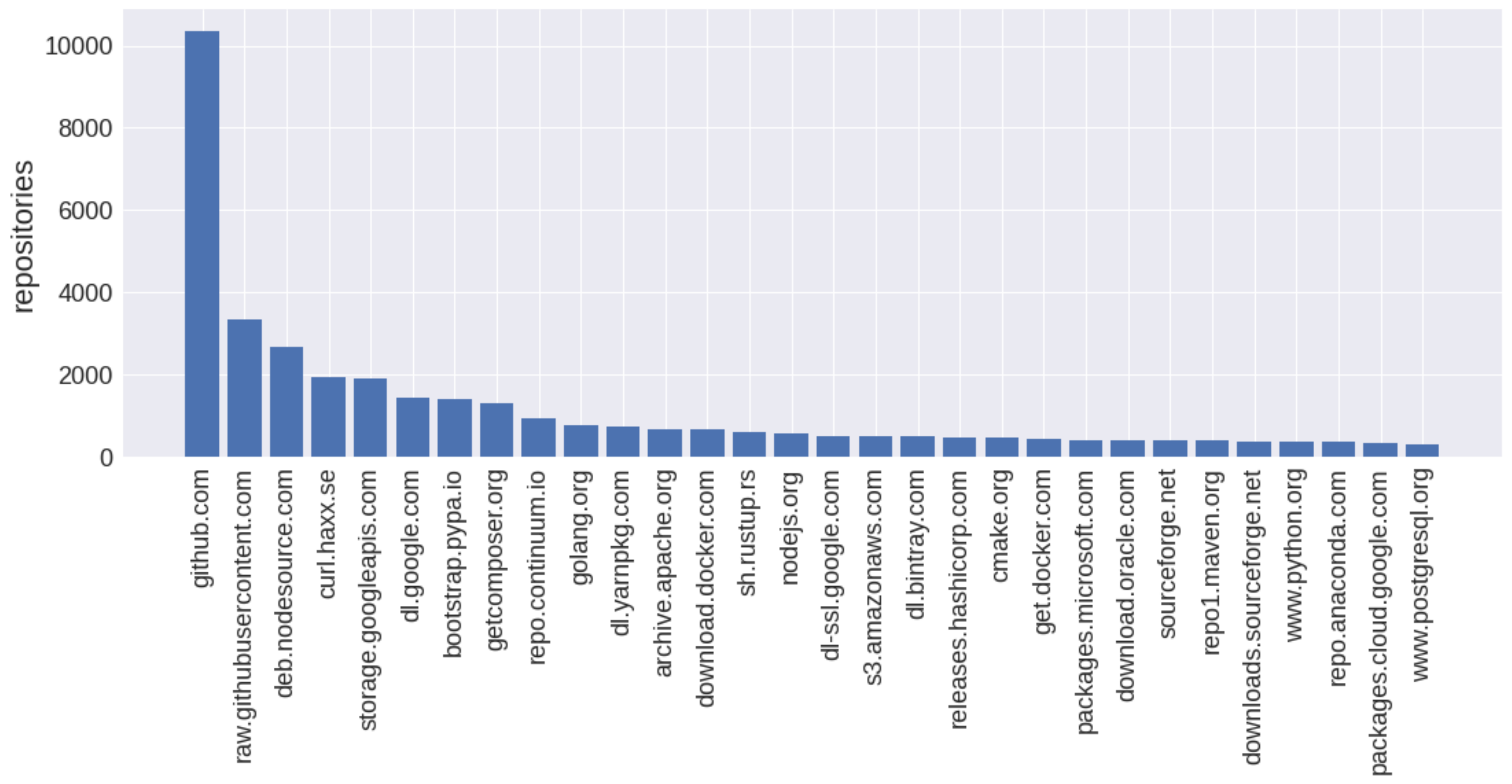}
	\caption{Top 30 domains of package sources.}
    \label{fig:domain}
\end{figure*}

\subsection{Extracting URLs}
\label{sec:url}
To investigate URL-based version-pinned packages (as mentioned in Section~\ref{sec:terminology}), we extract URLs specified in the arguments of \texttt{curl} and \texttt{wget} commands appearing in \texttt{ADD} and \texttt{RUN} commands in the Dockerfiles. In order to accurately extract URLs, a parsing of the Dockerfiles is performed in advance to create syntax trees.
Dockerfile has a hierarchical structure that can include shell code such as bash scripts in the arguments of \texttt{RUN} commands. To deal with this, the creation of the syntax tree is performed, similar to the previous work~\cite{Henkel:ICSE2020:20}. An example of parsing is shown in Figure~\ref{fig:parse}.
The following four steps are performed to extract URLs.
\begin{enumerate}
    \item A simple top-level parsing is performed. At this step, the argument of a \texttt{RUN} instruction is retained as a string literal. Given the example in Figure~\ref{fig:parse}a, we get the representation shown in Figure~\ref{fig:parse}b.
    \item A bash parser\footnote{\url{https://github.com/idank/bashlex}} is used to parse the arguments of \texttt{RUN} commands, which are held as string literals. From the representation in Figure~\ref{fig:parse}b, we obtain the representation in Figure~\ref{fig:parse}c.
    \item To improve readability and maintainability, version numbers in URLs are sometimes declared as variables in Dockerfiles. Therefore, the variable values in URLs are obtained from such variable definitions. The values are obtained from the default values in \texttt{ARG} commands and the variable declarations in the bash script in \texttt{RUN} commands, and replace the variables in the associated \texttt{ADD} and \texttt{RUN} commands.
    \item Strings in the \textit{source} of \texttt{ADD} commands that begin with \texttt{http} are obtained. Also, by referring to the created syntax trees, strings beginning with \textit{http} in the arguments of \texttt{curl} and \texttt{wget} commands in \texttt{RUN} commands are retrieved.
\end{enumerate}

We performed the above steps on all versions of 434,846 Dockerfiles obtained in Section~\ref{sec:obtaining-dockerfiles} to identify Dockerfiles that held URLs in at least one or more versions. As a result, there were 141,667 Dockerfiles in 27,620 repositories containing a total of 136,536 unique URLs.

Figure~\ref{fig:domain} shows the top 30 domains in the number of repositories in which the obtained URL appears. The \texttt{github.com} domain was the most common, with 10,362 repositories. This was followed by the \texttt{raw.githubusercon tent.com} domain with 3,357 repositories, and the \texttt{deb.nodesource .com} domain with 2,680 repositories. The total number of URLs retrieved consisted of 5,023 unique domains, suggesting that the Docker ecosystem leverages a variety of platforms as a source of remote files.
Since it is important to have repository groups that target the same package sets for meta-maintenance application, this research targets only the \texttt{github.com} domain found in about 10,000 repositories, which is overwhelmingly larger than the second and lower domains.

\subsection{Identifying Version-Pinned Packages}
Software packages and versions were identified from the extracted URLs. There are two patterns in the descriptions of version-pinned packages obtained from the GitHub.

The first pattern is the use of GitHub's release feature, which allows software packages to be distributed by git tags that point to specific versions. In this pattern, a package and a tag corresponding to a version can be obtained at the URL \texttt{https://github.com/:OWNER/ :REPO/releases/download/:TAG/:FILE}.

The second pattern is to obtain an archive of the source code; GitHub creates an archive of source code for each git tag that points to a specific version. In this pattern, a package and a tag corresponding to a version can be obtained at URLs \texttt{https://github.com/ :OWNER/:REPO/archive/refs/tags/:TAG(.zip|.tar.gz)} or\\ \texttt{https://github.com/:OWNER/:REPO/archive/:TAG(.zip|.\\tar.gz)}.

To check the validity of the \texttt{github.com} domain URLs collected in Section~\ref{sec:url}, we used the GitHub API v3 to obtain the release list\footnote{\url{https://docs.github.com/en/rest/releases/releases\#list-releases}} and tag list\footnote{\url{https://docs.github.com/en/rest/reference/repos\#list-repository-tags}} for each repository.
Since a release list contains all URLs in a repository, URLs that match the first pattern and are included in the release list are valid URLs. In addition, the tag part in the URL is checked against the tag list, and URLs that match the second pattern and whose tags are included in the tag list are valid URLs. These checks were performed on all versions of the analyzed Dockerfiles, and finally, 17,139 Dockerfiles and 7,914 repositories containing a total of 22,182 unique URLs were obtained. Packages were provided in 3,968 distinct repositories.

\section{Results}

This section presents the answers to our research questions.

\subsection{\RqOne}
\label{sec:rq1}

\begin{table}
\centering
\caption{Dockerfile classification regarding version-pinned packages}
\label{tab:classification}
\begin{tabular}{lrr}
\toprule
dormant & 4,016 & (23.4\%) \\
all packages deleted & 2,878 & (16.8\%) \\
packages with multiple versions & 60 & (0.4\%) \\
no package updated & 7,582 & (44.2\%) \\
package updated & 2,603 & (15.2\%) \\
\midrule
\textbf{sum} & \textbf{17,139} & \textbf{(100\%)} \\
\bottomrule
\end{tabular}
\end{table}

To understand how Dockerfiles are maintained, we classify Dockerfiles and repositories by their maintenance status.

\textbf{Classification of Dockerfiles.}
We classify Dockerfiles based on their statuses as follows.

\begin{itemize}
\item \textbf{dormant}: a Dockerfile is in a dormant (an unmaintained) repository. Similar to previous studies~\cite{Izeuierdo:MODELS2017,Coelho:WMO2017,Hata:ICSE2021}, we set one year as a threshold to consider dormant, that is, we consider a repository dormant or if the repository does not have commits after 2020.
\item \textbf{all packages deleted}: a Dockerfile does not contain packages in the latest commit because all packages had been removed at some point.
\item \textbf{packages with multiple versions}: The same package is used multiple times in the Dockerfile and their versions are different.
\item \textbf{no packages updated}: None of packages in the Dockerfile have been updated their pinned versions.
\item \textbf{packages updated}: The pinned package versions in the Dockerfile have been updated at least once.
\end{itemize}
Table~\ref{tab:classification} shows the frequencies of Dockerfile statuses.
We observed that the majority of Dockerfiles were ``no package updated'', but 15.2\% of Dockerfiles were maintained with ``package updated''.

\begin{table}
\centering
\caption{Repository classification based on their Dockerfiles containing version-pinned packages}
\label{tab:type}
\begin{tabular}{lrr}
\toprule
with update & 1,454 & (18\%) \\
no update & 2,890 & (37\%) \\
other & 3,570 & (45\%) \\
\midrule
\textbf{sum} & \textbf{7,914} & \textbf{(100\%)} \\
\bottomrule
\end{tabular}
\end{table}

\textbf{Classification of repositories.}
Table~\ref{tab:type} summarizes the frequencies of repository types as described below.

\begin{itemize}
\item \textbf{with update}: A repository contains at least one ``package updated'' Dockerfile.
\item \textbf{no update}: A repository contains only ``no packages updated'' Dockerfiles.
\item \textbf{other}: A repository contains only ``dormant'', ``all packages deleted'', or ``packages with multiple versions'', making it difficult to apply the meta-maintenance approach analyzed in this study.
\end{itemize}

The repositories of ``with update'' are expected to help identify useful package updates, and the repositories of ``no update'' could be targeted for update assistance. A total of 4,344 repositories will be used to answer the following research questions.

\begin{tcolorbox}[title=Summary]
Of the repositories that had contained Dockerfiles with version-pinned packages, 18\% had version updates, 37\% had no version updates, and the remaining 45\% are dormant repositories or the Dockerfile with version-pinned package was gone.
\end{tcolorbox}

\subsection{\RqTwo}

From all versions of each of the Dockerfile in the 4,344 repositories identified in Section~\ref{sec:rq1}, we identify combinations of version-pinned packages that appear in two or more different repositories as package sets.

\begin{figure}
    \centering
    \begin{subfigure}{\columnwidth}
        \centering
        \includegraphics[width=\linewidth]{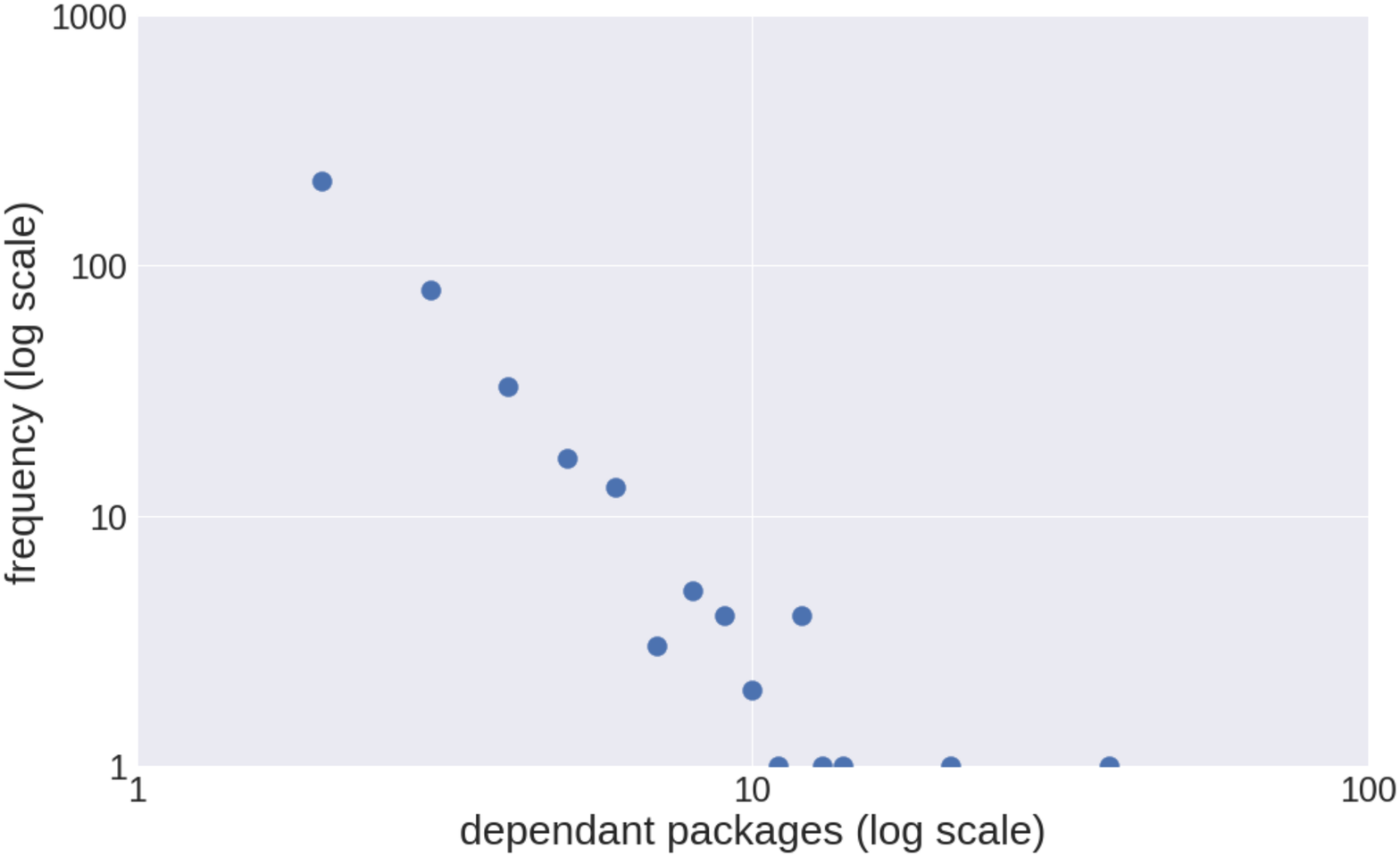}
		\caption{Frequency of package set sizes.}
	    \label{fig:size}
    \end{subfigure}
    \begin{subfigure}{\columnwidth}
        \centering
        \includegraphics[width=\linewidth]{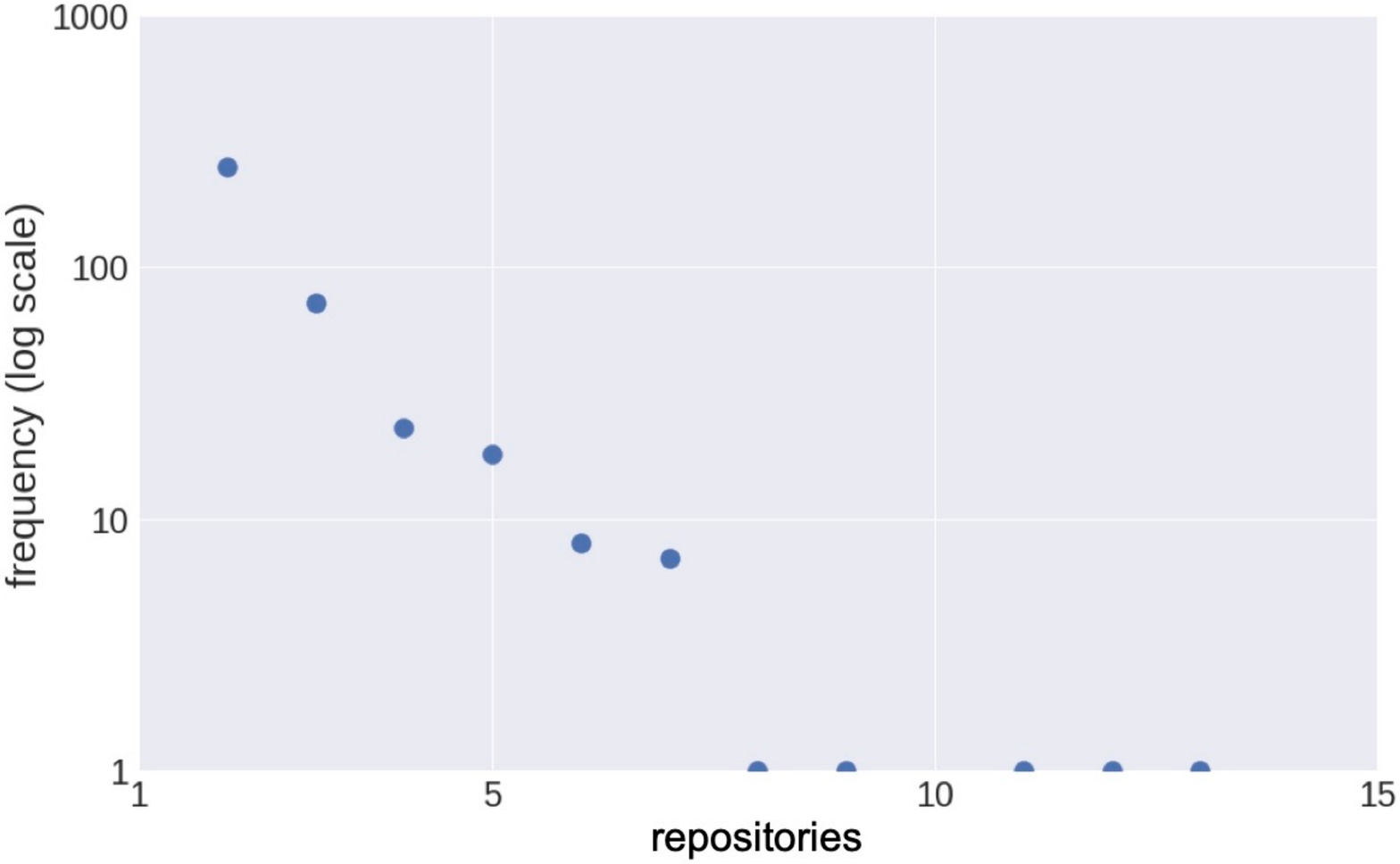}
		\caption{Frequency of repo group sizes.}
	    \label{fig:repo}
    \end{subfigure}
    \caption{Distributions of the number of packages and the number of repositories with same package sets.}
    \label{fig:dist}
\end{figure}

The number of packages that comprise a package set (package set size) and their frequency are shown in Figure~\ref{fig:size}. Package sets consisted of a minimum of two and a maximum of 38 packages. Package sets consisting of two packages were found in a total of 219 repositories. The number of repositories using package sets consisting of nine or more packages was less than four in all groups.

Figure~\ref{fig:repo} shows the frequency of the number of repositories in a group of repositories with Dockerfiles containing the same package set. The repository groups were composed of a minimum of two and a maximum of 13 repositories; the repository group composed of two repositories had the highest frequency of 252. The frequency decreased as the number of repositories increased, with a frequency of seven for repository groups consisting of seven repositories and one for all repository groups consisting of eight or more repositories. A total of 385 repository groups were identified as having common package sets. 

\begin{table}
\centering
\caption{Packages per programming languages}
\label{tab:lang}
\begin{tabular}{lrl}
\toprule
language & \# & package repository examples \\
\midrule
C & 88 & samtools/samtools, alexdobin/STAR \\ && samtools/htslib, arq5x/bedtools2 \\
Go & 85 & containernetworking/plugins \\ && docker/compose, containerd/containerd \\
C++ & 80 & opencv/opencv, mstorsjo/fdk-aac \\ && Intel-Media-SDK/MediaSDK \\
Java & 10 & broadinstitute/picard, bazelbuild/bazel \\
Shell & 9 & sgerrand/alpine-pkg-glibc \\ && just-containers/s6-overlay \\
JavaScript & 8 & novnc/noVNC, microsoft/DLWorkspace \\ && dustinblackman/phantomized \\
Python & 7 & Yelp/dumb-init, Netflix/vmaf \\ && novnc/websockify \\
PHP & 5 & squizlabs/PHP\_CodeSniffer \\
Haskell & 4 & jgm/pandoc \\
Erlang & 4 & erlang/otp \\
HTML & 4 & OSGeo/libgeotiff \\
Perl & 4 & trinityrnaseq/trinityrnaseq \\
TypeScript & 3 & cdr/code-server \\
Scala & 3 & lampepfl/dotty \\
Fortran & 3 & xianyi/OpenBLAS \\
PLpgSQL & 1 & darold/pgFormatter \\
Elixir & 1 & elixir-lang/elixir \\
OCaml & 1 & ocaml/opam \\
Assembly & 1 & videolan/x265 \\
WebAssembly & 1 & WebAssembly/binaryen \\
Rust & 1 & sharkdp/bat \\
C\# & 1 & GitTools/GitVersion \\
Kotlin & 1 & neo4j-graphql/neo4j-graphql \\
Crystal & 1 & crystal-lang/crystal \\
SMT & 1 & Boolector/boolector \\
D & 1 & lldc-developers/ldc \\
Makefile & 1 & rabbitmq/rabbitmq-server \\
V & 1 & vlang/v \\
R & 1 & kundajelab/phantompeakqualtools \\
Cmake & 1 & googleapis/cpp-cmakefiles \\
\bottomrule
\end{tabular}
\end{table}

Table~\ref{tab:lang} summarizes the programming languages of the repositories that provide the packages included in the identified 385 package sets. The programming languages of the repositories were identified using the GitHub API v3. The largest number of package repositories was written in C, such as samtools, a DNA data analysis tool. Go was the next most frequent language, followed by 
C++ language. 
C, Go, and C++ were the top three languages of package repositories frequently appear in the identified package sets.

\begin{tcolorbox}[title=Summary]
The identified package sets consisted of a minimum of two and a maximum of 38 packages, for a total of 385 sets. Most of the packages were written in C, Go, and C++.
\end{tcolorbox}

\subsection{\RqThree}

To understand how Dockerfiles with the same package set evolve differently, we classify package sets by their maintenance status and investigate version differences within Dockerfiles with common package sets.

\textbf{Classification of package sets.}
Considering the combination of package versions in a package set, we classify repository groups by Dockerfile with the same package set.
For a set of packages $a, b, ...$, a particular version combination in a single Dockerfile is denoted as $(a_i, b_j, ...)$, where $i$, $j$ are versions.
We consider the mathematical \textit{product order} defined as $(a_i, b_j) \leq (a_p, b_q)$ if and only if $a_i \leq a_p$ and $b_j \leq b_q$, that is, $a_p$ and $b_q$ are newer versions of packages $a$ and $b$, compared to $a_i$ and $b_j$.

\begin{itemize}
\item \textbf{no update}: A repository group consists only of ``no update'' repositories, that is, the Dockerfiles of all repositories belonging to this group have not updated version-pinned packages.
\item \textbf{equivalent}: The version combinations of all packages in the package set of the latest commit of the Dockerfile is identical in all repositories. This repository group contains at least one repository ``with update''.
\item \textbf{incomparable}: The version combinations of all packages in the package set is not the product order, that is, there are version combinations $(x_i, y_j)$ and $(x_p, y_q)$ where $x_i < x_p$ and $y_j > y_q$.
\item \textbf{comparable}: The version combinations of all packages in the package set is the product order and not all version combinations are identical.
\end{itemize}

\begin{table}
\centering
\caption{Package set classification regarding version-pinned packages}
\label{tab:set}
\begin{tabular}{lrr}
\toprule
no update & 45 & (12\%) \\
equivalent & 36 & (9\%) \\
incomparable & 96 & (25\%) \\
comparable & 208 & (54\%) \\
\midrule
\textbf{sum} & \textbf{385} & \textbf{(100\%)} \\
\bottomrule
\end{tabular}
\end{table}

Table~\ref{tab:set} summarizes the frequencies of package sets.
Package sets of ``no update'' and ``equivalent'' is not meta-maintenance applicable because there are no useful package updates.
Package set of ``incomparable'' might contain useful package updates, but there is no combination of all versions being equal or newer than the others, making it difficult to identify useful package updates and apply them to other Dockerfiles.
Since ``comparable'' package sets are clearly considered meta-maintenance applicable, only these package sets will be analyzed hereafter. More than half of the package sets in the analysis were ``comparable'' package sets, with 363 repositories containing these package sets.

\begin{figure}
    \centering
    \begin{subfigure}{\columnwidth}
        \centering
        \includegraphics[width=\linewidth]{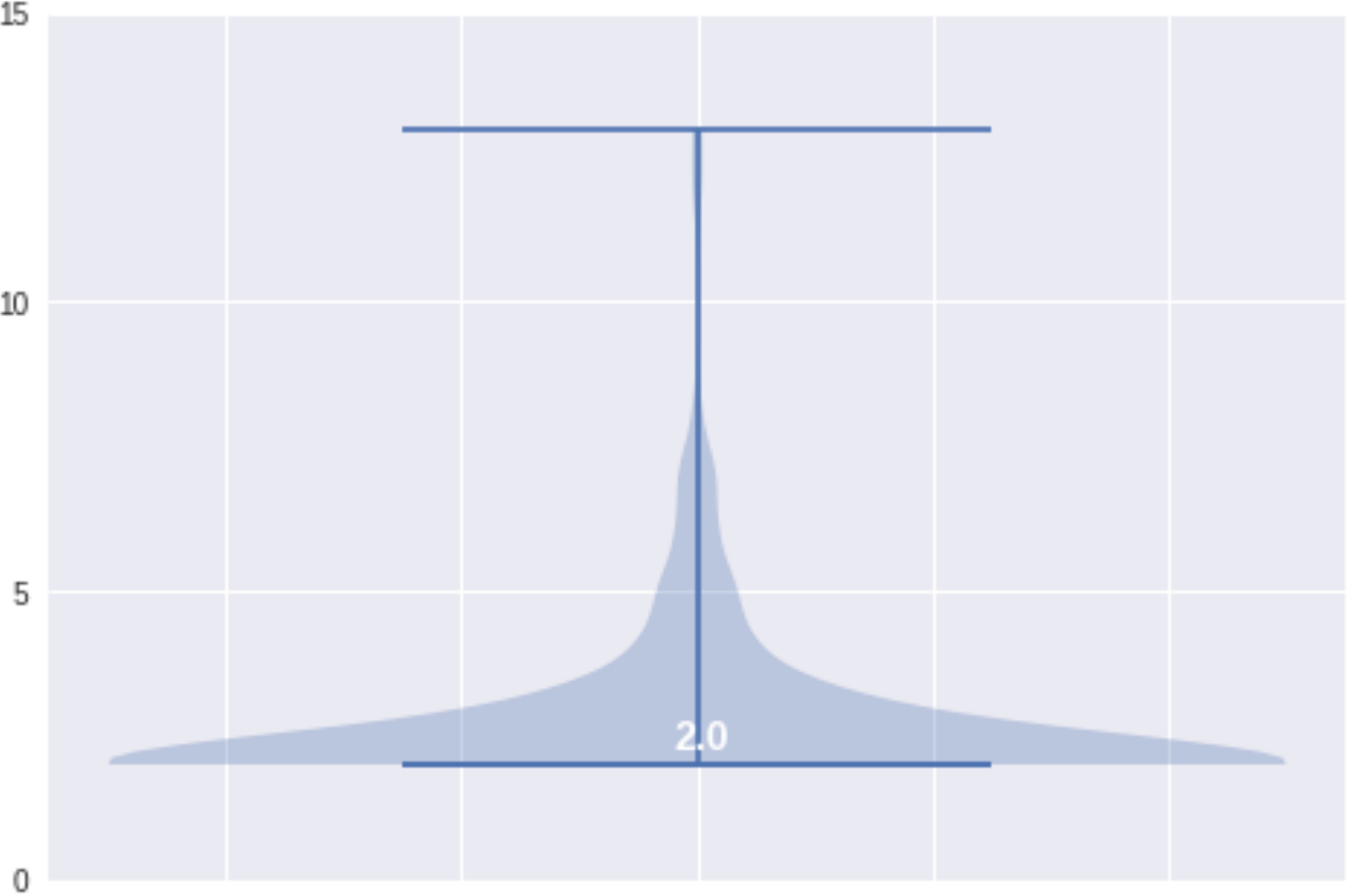}
		\caption{Number of repositories with differences.}
	    \label{fig:different}
    \end{subfigure}
    \begin{subfigure}{\columnwidth}
        \centering
        \includegraphics[width=\linewidth]{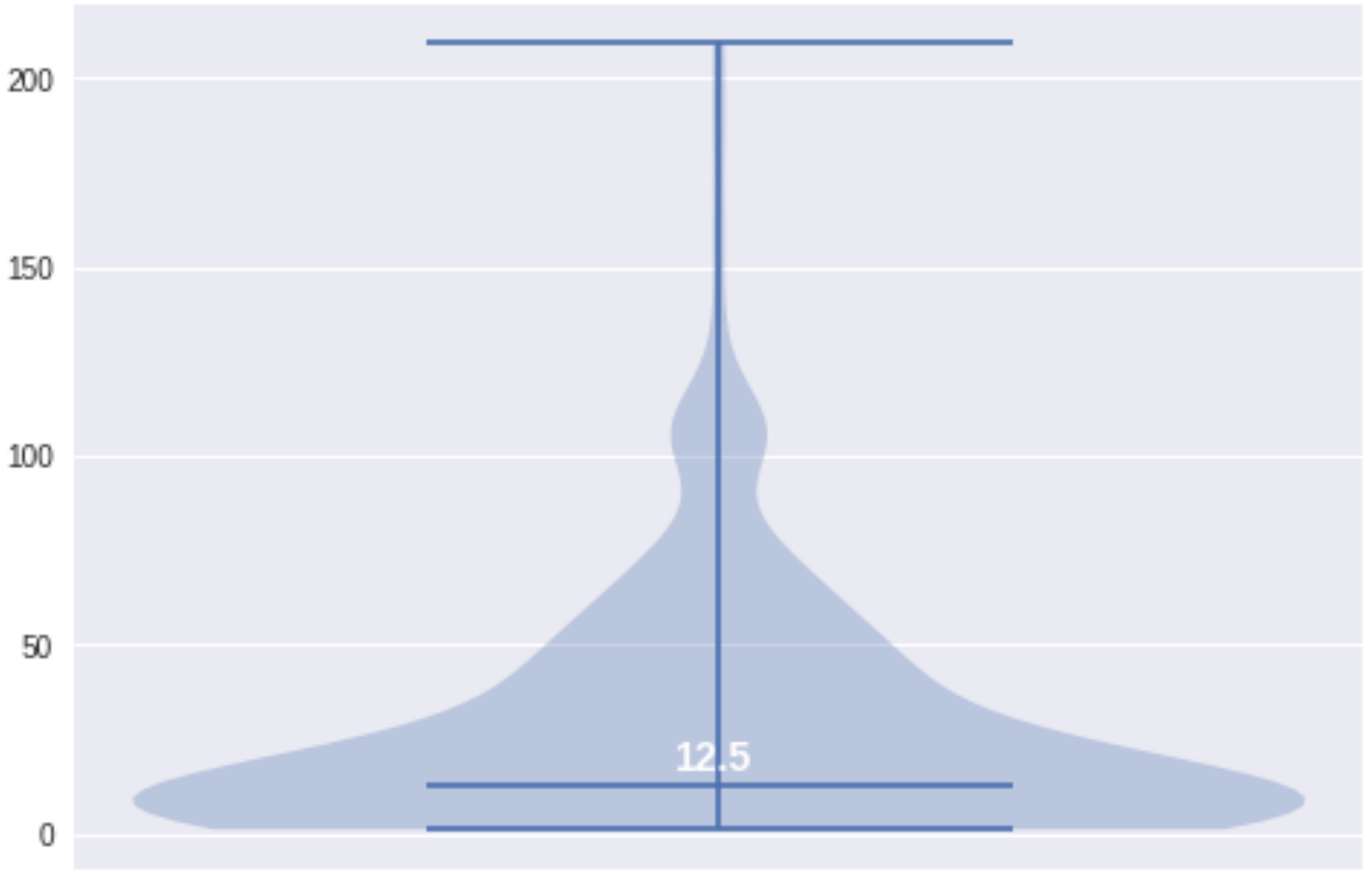}
		\caption{Maximum version difference.}
	    \label{fig:biggest}
    \end{subfigure}
    \caption{Version differences in package sets.}
    \label{fig:evol}
\end{figure}

\textbf{Version differences within a package set.}
In order to understand how potentially meta-maintenance approach can support package version updates, we investigate version differences in each package set used in different repositories.
From the ``comparable'' package sets identified above, we counted the number of Dockerfiles that exist in different repositories, including packages of different versions. As shown in Figure~\ref{fig:different}, the median value was two. 

Then we investigate how much version difference there is between a Dockerfile using a newer version of a package and a Dockerfile using an older version of the same package in repositories that use the same package set.
The version difference is measured by how many differences there are between the git tags corresponding to the two versions. For example, suppose a package has git tags set in the order v1.0, v1.1, v1.2, v1.3, v2.0. If git tags corresponding to versions to be measured are v1.0 and v2.0, the version difference is four as there is a difference of four tags.
In the 208 ``comparable'' package sets, the distribution of the largest version differences for all packages in the package set is shown in Figure~\ref{fig:biggest}. The median was 12.5. 

\begin{tcolorbox}[title=Summary]
Repository groups to which more than half of the package sets belonged contained Dockerfiles with newer version combinations compared to the version combinations in other Dockerfiles.
The median value for that package set was 12.5 
for the largest version difference in a package set.
\end{tcolorbox}

\subsection{\RqFour}

To understand the potential of meta-maintenance, we perform a qualitative investigation by submitted pull requests with the following message to update multiple version-pinned packages.
\begin{quote}
    \textit{`In this pull request, I am updating <<package from XX to XY and package from WW to WZ. Since these updates are being done in ..., I’m wondering if this project can update the packages as well.'}
\end{quote}
Note that for each pull request, the first author must manually update the packages and perform the build.
As a result, each pull request was submitted to four GitHub projects, out of the four, one received no response, 
another failed some bot checks. 
We did however received some developer feedback from two cases, both which are discussed below.

\textbf{Case 1: Bumping libgit2, ligssh2 packages for the Gitcovex dockerfiles.}
We sent a pull request to the Gitconvex, 
which is a Web application for managing git repositories. 
Specifically, we asked them to update the libraries, libgit2 from 1.1.0 to 1.1.1 and libssh2 from 1.9.0 to 1.10.0.
Furthermore, we informed them that libssh2@v1.9.0 was affected by a vulnerability CVE-2019-17498.\footnote{\url{https://nvd.nist.gov/vuln/detail/CVE-2019-17498}}
Interestingly, developers responded by saying that:
\begin{quote}
    \textit{We have updated the docker file in the backend repo for gitconvex Dockerfile but due to lack of bandwidth, the maintainer couldn't test and merge it.}
\end{quote}
This answer reveals use case where many of these updates may require additional resources that the maintainer might not have. 
This indicates that updating packages in Dockerfiles is not an easy task, even with the recommended version combinations.

\textbf{Case 2: Bumping libde265, libheif packages for Bref.}
\textsf{Bref} is a software project that has the aim to make running PHP applications simple by using serverless technologies.
The affected repository is the PHP extensions.\footnote{\url{https://github.com/brefphp/extra-php-extensions}}
Our submitted pull request sparked much discussion, that at first started with asking what was our motivation to update the library:
\begin{quote}
    \textit{ What would the benefits for imagick be when we update these dependencies?}
\end{quote}
After providing a general response to the benefits of updating from a maintenance perspective, the developers then explained their hesitancy to update, stating that:
\begin{quote}
    \textit{
True, but so is stable software. Each change, big or small, is a risk for bugs and issues. If a change does not bring any value it should not be considered.}
\end{quote}
Furthermore, the developer is careful to highlight that they do see the importance of updating.
\begin{quote}
    \textit{Note, Im not against this change. Im just curious why this is needed? How will it help you when you run imagick on with Bref? Will it introduce a new feature? Will it fix a bug for you? Or will it help you in any other way?}
\end{quote}
During communication with the developers, we found a security problem (CVE-2021-35452) in the new version of the proposed package. We informed them of this and suspended the update recommendation.
It suggests that security-related updates should be more cautious.

Although these are single cases, we believe that they highlight some of the reasoning behind the decision to update. 
Developers are less likely to update when there is no tangible advantages, and would rather having stable software.

\begin{tcolorbox}[title=Summary]
We learned that developers may require additional motivation, may it be tangible advantages, before they consider updates. 
Consistent with prior work, updating although may be trivial, but may require additional resources and risks instable software.
\end{tcolorbox}

\section{Discussion}

Based on our results, we now summarize the barriers that need to be addressed.
Then we discuss the limitations of our study.

\subsection{Implications}
This paper explores the state-of-the-practice for how meta maintenance is leveraged for updating multiple version-pinned packages.
Here are the key implications that are takeaways from the study:

\begin{itemize}
    \item \textit{Motivation with foreseeable benefits are needed to convince developers to update}.
    We show that developers will only consider the proposed package set updates, if there are realized benefits behind the update.
    As shown in \textbf{RQ4}, developers may lack the resources and efforts to execute such updates. Tool builders should also consider this aspect when presenting developers with automated approaches.
   
    \item \textit{The effects of the dormant packages is unknown}. The findings from \textbf{RQ1} reveal that almost 45\% are dormant.
    Since this makes meta-maintenance difficult, there should be better ways to understand the effects of being outdated. 
    This is regarded as future work. The implication is that these repositories miss out on the benefits of applying meta-maintenance.
    
    \item \textit{Updates come in package sets}.
    Results from \textbf{RQ2} and \textbf{RQ3} show that package updates are usually bundled in package sets.
    The implication for developers and for tool builders is to understand that recommended updates are not singular, and that combinations of updates would be preferable.

    \item \textit{Leverage advanced outcomes from credible projects to increase software supply chain security in software ecosystems}.
    The Alpha-Omega Project appears to support version updates for specific projects. Dockerfile meta-maintenance might be more practical if the proposed version combinations could be continuously collected from credible projects with reliable update information.
\end{itemize}

\subsection{Threats to Validity}
Threats to \textit{construct validity} exist in our data collection procedure.
Our criteria for selecting repositories may have ignored recent active projects and less popular projects.
However, we have detailed our data collection process (See Section \ref{sec:obtaining-dockerfiles}) so that the process for collection can be replicated and improved in the future. 
Threats to \textit{external validity} exist in our generalization of our case study of GitHub repositories to other OSS projects that existing in other platforms, even those that are closed.
Another limitation of this study is that only URL-based package acquisitions in the GitHub domain were included in the analysis.
We acknowledge these threats and look for future work to test this generalization in the future.
To mitigate threats to \textit{reliability}, we will prepare an online appendix of our studied dataset with associated information. 

\section{Related Work}
In this section, we discuss some key related studies.

\textbf{Empirical studies in Docker ecosystems.}
These studies provide analyses of the Docker ecosystem from different aspects.
Docker Hub provides a large number of Docker images for each software system, which contains a variety of installed libraries and versions \citep{Ibrahim:EMSE2020:8}.
\citet{Haque:ESEM2020:22} showed that the Docker community has a smaller number of experts compared to other communities.
The evolution of Dockerfile affects the containerization outcomes of the project \citep{Zhang:COMPSAC2018:3}.
According to the study of \citet{Cito:MSR2017:1}, Dockerfiles are not changed often as they are around 3.81 revisions per year.
\citet{Wu:APSEC2020:10} found that Dockerfiles usually changed with other files.
They also suggested that there is an association between the frequency of changes in Dockerfiles and the popularity of projects.
Several studies show that there are cloned Dockerfiles in the software project \citep{Oumaziz:ICSME2019:5, Eng:MSR2021:12, Zhao:CLUSTER2019:6}.
From investigated by \citet{Oumaziz:ICSME2019:5}, developers maintained multiple Dockerfiles as they want to support multiple working environments at the same time.
In this case, the maintainers have to make the identical changes to those multiple Dockerfiles.
Several studies focus on the quality of Dockerfile.
\citet{Wu:MSR2020:11} found that up to 85.2\% of Docker projects have at least one broken build.
They also found that the more frequently developers build projects, the less likely projects fail to build, while the more often developers fail to build, the longer it takes to fix issues.
\citet{Cito:MSR2017:1} found that 28.6\% of quality issues within Dockerfile arise from missing version pinning.
\citet{Lin:ICSME2020:9, Yin:APSEC2018:15} investigated Docker image tagging practices.
They found that the usage of obsolete images (e.g., Ubuntu base image) is increasing, which may pose security risks for users.
They also find that there are issues with the usage of the latest tag.
Currently, the support for developers to find the Docker images is limited as users can only search them by name \citep{Brogi:IC2E2017:13}.
\citet{Wu:IEEEA2020:7} investigated the smell of Dockerfile code within GitHub projects and found that nearly 84\% of them have smells with 62\% of Dockerfile instructions being affected by those smells.
Additionally, \citet{Lu:IEEEA2019:4} reported that the temporary file smells in Dockerfile widely exist.
Even though the smells existed in Dockerfile, the trend is decreasing over time \citep{Lin:ICSME2020:9, Eng:MSR2021:12}.
\citet{Henkel:ICSE2020:20} found that Dockerfiles on GitHub in the wild is nearly five times worse compared to ones by experts.
\citet{Zhang:COMPSAC2018:3} suggested that using more diverse instructions, having fewer image layers, and adopting an official image can reduce Dockerfile quality issues.
In this paper, we focused on the package version maintenance issue in Docker images by conducting an empirical experiment on package usage and evolution in Dockerfiles.

\textbf{Updating and vulnerability issues in Docker ecosystems.}
Cybercrime is a growing threat worldwide as it is expected to cost damages around 10.5 trillion dollars annually by 2025 \citep{Web:Cybercrime}.
One of the cyber-attack is to exploit software vulnerabilities.
Several studies explored the impact of vulnerabilities within different software ecosystems through the dependency analysis \citep{Kikas:MSR2017,Linares:MSR2017,Lauinger:NDSS2016}, while some studies focused on the Docker community \citep{Shu:CODASPY2017:24, Wist:SAM2020, Ibrahim:EMSE2020:8}.
As analyzed by \citet{Ibrahim:EMSE2020:8}, the official Docker images are not always the most secure images for developers.
\citet{Shu:CODASPY2017:24} found that Docker images have an average of 180 or more vulnerabilities in all versions.
They also found that lots of Docker images have not been updated for over 100 days.
\citet{Wist:SAM2020} analyzed the vulnerabilities from 2,500 Docker images and found that only 17.8\% of Docker images were completely free from vulnerabilities.
To eliminate the risk of vulnerabilities, developers have to frequently update their software artifacts \citep{Dissanayake:IST2022}.
However, multiple researchers reported that there are lags in updating those artifacts which caused the supply chain remains vulnerable \citep{Shu:CODASPY2017:24, Chinthanet:EMSE2021, Gonzalez:OSS2017, Zerouali:EMSE2021, Ihara:OSS2017, Kula:EMSE2017}.
In the case of Docker images, \citet{Liu:ESORICS2020} found that the patch for vulnerabilities was significantly delayed (i.e., 422 days on average).
Keeping a lot of outdated packages could lead to vulnerability prone in the images.
\citet{Zerouali:EMSE2021} analyzed the Debian-based Docker images and found that the number of old packages within the images correlated with the number of vulnerabilities affected those images, while there is no correlation between the number of old packages and the number of bugs of those images.
They also find that the official image had fewer lags compared to the community image in terms of package lag, time lag, version lag, vulnerability lag, and bug lag.
In this paper, we explored the applicability of the meta-maintenance approach in order to update multiple version-pinned packages in Dockerfiles, which we found that it is applicable.

\textbf{Tools for Dockerfile maintenance.}
There are several tools available to assist developers in maintaining the Dockerfile.
RUDSEA analyzes the environmental requirements to suggest whether developers have to update their Dockerfile or not \citep{Hassan:ASE2018:14}.
Binnacle detects any violations of 15 best practices within Dockerfile \citep{Henkel:ICSE2020:20}.
Shipwright semi-automatically corrects Dockerfile \citep{Henkel:ICSE2021:25}.
Dockerfile Linter performs a static analysis on Dockerfile and points out convention violations \citep{Web:DockerfileLinter}.
Dependabot and Renovate create a pull request (PR) for repositories on GitHub to update dependencies and fix vulnerabilities \citep{Web:Dependabot, Web:renovate}.
In this paper, we suggested that a tool to help developers find specific changes and maintain code could be practically useful.

\section{Conclusion}
To understand the applicability of meta-maintenance for updating multiple version-pinned packages in Dockerfiles, we conducted an exploratory study of 7,914 projects with 17,139 Dockerfiles on GitHub to determine
(i) the maintenance status of Dockerfiles with version-pinned packages; 
(ii) the characteristics of package sets used in multiple Dockerfiles; 
(iii) the evolution and maintenance aspects of the package set; and
(iv) developers' perceptions of recommended updates.
We found 385 repository groups with the same package sets, and 208 groups had Dockerfiles with newer version combinations compared to others, which could be useful in meta-maintenance. Our findings support the applicability of meta-maintenance to update multiple version-pinned packages. There are also promising avenues and challenges for future research, such as investigating relationships among package sets, developing techniques to identify useful changes, and supporting tools to automate meta-maintenance.

\section*{Data Availability}

Our online appendix (\url{https://doi.org/10.5281/zenodo.6523208}) containing the lists of 7,914 repositories,
17,139 dockerfiles,
335 packages, and
the details of 208 comparable package sets appearing in 363 repositories.


\bibliographystyle{ACM-Reference-Format}
\bibliography{base}


\end{document}